\documentclass[usegraphicx,usenatbib]{mn2e}
\usepackage{times}
\renewcommand\[{\begin{equation}}
\renewcommand\]{\end{equation}}
\def\parn{\par\noindent}

\catcode`\@=11
\def\gsim{\ifmmode{\mathrel{\mathpalette\@versim>}}
    \else{$\mathrel{\mathpalette\@versim>}$}\fi}
\def\lsim{\ifmmode{\mathrel{\mathpalette\@versim<}}
    \else{$\mathrel{\mathpalette\@versim<}$}\fi}
\def\@versim#1#2{\lower 2.9truept \vbox{\baselineskip 0pt \lineskip
    0.5truept \ialign{$\m@th#1\hfil##\hfil$\crcr#2\crcr\sim\crcr}}}
\catcode`\@=12

\def\eps{{\cal E}}

\def\fid{f_{\rm i}}

\def\ln{\hbox{${\rm ln}\, $}}

\def\parn{\par\noindent}

\def\psit{\Psi_{\rm T}}

\def\ra{r_{\rm a}}
\def\rai{r_{\rm ai}}
\def\rc{r_{\rm c}}

\def\roi{\rho _{\rm i}}

\def\sa{s_{\rm a}}

\def\sigr{\sigma_{\rm r}}
\def\sigt{\sigma_{\rm t}}

   \title[Consistency of generalized Cuddeford systems]
   {Consistency criteria for generalized Cuddeford systems}

   \author[Ciotti \& Morganti]
           {Luca Ciotti \& Lucia Morganti\thanks{Current address:
 Max-Planck-Institut f\"ur Ex. Physik, Giessenbachstra\ss{}e,
 D-85741 Garching, Germany}\\Astronomy Department, 
      University of Bologna, via Ranzani 1, 40127 Bologna, Italy}

\date{Accepted 2009 September 11.  Received 2009 September 10; in original form 2009 July 9}
\pubyear{2009}

\begin{document} 
\maketitle

\begin{abstract} 
General criteria to check the positivity of the distribution function
(phase--space consistency) of stellar systems
of assigned density and anisotropy profile
are useful starting points in Jeans--based modeling.
Here we substantially extend previous results,
and we present the inversion formula and the analytical necessary and sufficient conditions 
for phase--space consistency of the family
of multi--component Cuddeford spherical systems:
the distribution function of each density component of these systems 
is defined as the sum of an arbitrary number of Cuddeford distribution functions with arbitrary values
of the anisotropy radius, but identical angular momentum exponent.
The radial trend of anisotropy that can be realized by these models is therefore very general.
As a surprising by--product of our study,
we found that the ``central cusp--anisotropy theorem'' 
(a necessary condition for consistency relating 
the values of the central density slope 
and of the anisotropy parameter)
holds not only at the center, but at all radii
in consistent multi--component generalized Cuddeford systems. 
This last result suggests that the so--called 
mass--anisotropy degeneracy could be less severe
than what is sometimes feared.
\end{abstract}


\section{Introduction}
In the study of stellar systems
based on the ``$\rho$--to--$f$'' approach
(where $\rho$ is the material density and $f$
is the associated phase--space distribution function,
hereafter DF; e.g. see Bertin 2000, Binney \& Tremaine 2008),
the density distribution is given,
and specific assumptions on the internal dynamics of the model are made.
In some special cases inversion formulae exist so that the DF 
can be obtained, usually in integral form or as series expansion (see, e.g.,
Fricke 1952; Lynden--Bell 1962; Osipkov 1979; Merritt 1985; 
Dejonghe 1986, 1987; Cuddeford 1991; Hunter \& Qian 1993; Ciotti
\& Bertin 2005). 
Once the DF of the system is derived,
a non--negativity check is (or should be) performed,
and in case of negative values the model must be discarded as unphysical.
Indeed, a minimal but essential requirement to be met by the DF
(of each component) of a stellar dynamical model 
is positivity over the accessible phase--space.
This requirement, the so--called phase--space consistency,
is much weaker than the model 
stability, but it is stronger than the fact that the Jeans equations have a
physically acceptable solution.
However, the difficulties inherent in the operation of
recovering analytically the DF prevent in general a simple consistency analysis,
and numerical inspection of the inversion integral
is required.
As a consequence, the reasons underlying
consistency or inconsistency of a proposed model
are somewhat obscured by the numerical nature of the solution. 
Fortunately, criteria for phase--space consistency 
that can be applied without an explicit recovery of the DF 
are known and widely used.
For example, analytical necessary and sufficient conditions
for consistency of multi--component systems
with Osipkov--Merritt anisotropy
(Osipkov 1979, Merritt 1985, hereafter OM) were derived 
in Ciotti \& Pellegrini (1992,
hereafter CP92; see also Tremaine et al. 1994)
and applied in several investigations
(e.g., Ciotti 1996, hereafter C96; Ciotti 1999, hereafter C99; 
Ciotti \& Lanzoni 1997; Ciotti \& Morganti 2009,
hereafter CM09; Ciotti, Morganti \& de Zeeuw 2009).
Such conditions revealed not only simple and useful to investigate
the phase--space consistency of OM models,
but also helpful to elucidate the different roles of total potential, orbital anisotropy,
and stellar and dark matter density profiles
in making a model unphysical.

More recently, the ``central cusp--anisotropy 
theorem'' (An \& Evans 2006, hereafter AE06),
a necessary condition for consistency relating 
the values of the central density slope 
and of the anisotropy parameter $\beta$ 
(see equation~[\ref{betaOM}]) has been proved\footnote{The same inequality 
was also reported in equation~(28) in de Bruijne et al. (1996).}. 
This condition was derived for constant anisotropy systems,
and then generalized asymptotically 
to the central regions of spherical systems
with arbitrary anisotropy distribution.
A remarkable property of the density slope--anisotropy inequality
is that it actually holds rigorously at \textit{every}
radius in constant anisotropy systems,
and not only at their center (AE06).
Surprisingly, in CM09 we showed that the CP92 necessary condition 
for model consistency can be formally rewritten as
the AE06 inequality, that consequently holds
at each radius not only in constant anisotropy systems,
but also in multi--component OM systems!
This curious result prompted us to investigate the phase--space consistency
of Cuddeford (1991) anisotropic systems,
as they generalize both constant and OM anisotropy
and an explicit inversion formula exists,
so that necessary and sufficient conditions for consistency 
can hopefully be found, extending those of CP92.
In addition, Cuddeford anisotropy allows to explore systems
in which the central regions may be tangentially anisotropic,
at variance with the OM cases.

Actually, we found it possible to extend our study to the very general case
of multi--component, generalized Cuddeford systems,
i.e. spherical systems in which the DF of \textit{each} distinct
density component is assumed to be the sum of an arbitrary number
of Cuddeford DFs with arbitrarily different anisotropy radii,
but identical angular momentum exponent
(see equation~[\ref{sumCud}]).
In this paper we show how the family of necessary and sufficient conditions 
for model consistency can be derived for generalized Cuddeford anisotropic systems.
We also found that the first of the necessary conditions
coincides again with the density slope--anisotropy theorem,
thus demonstrating that such inequality must be satisfied at all radii
also in the whole family of consistent, multi--component
generalized Cuddeford systems.

The paper is organized as follows.
In Section 2 we recall the fundamental properties
of OM and constant anisotropy systems,
and the associated consistency criteria.
In Section 3 we derive the family of consistency criteria 
for the larger class of multi--component 
galaxy models with generalized Cuddeford anisotropy.
Then, in Section 4 some illustrative applications 
of the new phase--space consistency criteria
are presented, and in Section 5
the main results are summarized, with a brief discussion
of the relevance of the new findings 
for the mass--anisotropy degeneracy problem.
In the Appendix we prove that the first of the necessary conditions
for phase--space consistency of multi--component
generalized Cuddeford systems can be rewritten as the
density slope-anisotropy inequality,
that must hold at all radii.

\section{Consistency criteria for multi--component Osipkov--Merritt systems}
In this Section we summarize the main features
of the OM inversion procedure, focusing
on the arguments upon which the derivation of the CP92 
necessary and sufficient conditions for phase--space consistency is based:
in fact, similar arguments will be applied 
to generalized Cuddeford systems in Section 3.

To fix the notation, we say that a multi--component stellar system 
described by a sum of different density components $\roi$ is called consistent 
if each DF $\fid$ is non--negative over the whole accessible phase--space.
However, as all the conditions presented in this paper
hold for each $\roi$, for simplicity
from now on the index $i$ is not indicated,
except when required for clarity.

The OM prescription assumes that the DF
supporting each density component
depends on the energy and on the angular momentum modulus 
of stellar orbits as
\begin{equation}\label{fOM}
f=f(Q),\quad Q=\eps-\frac{J^2}{2\ra^2},
\end{equation}
and $f(Q)=0$ for $Q\leq0$.
In the formula above $\eps=\psit-v^2/2$ is the binding energy per unit mass,
$\psit=-\Phi_{\rm T}$, where $\Phi_{\rm T}$ is the potential
due to the combined effect of all the components $\roi$,
and $\ra$ is the so--called anisotropy radius of each component
(e.g. see Binney \& Tremaine 2008).
Each density component of a multi--component OM system 
is characterized by a DF of the family~(\ref{fOM}),
in general with different $\rai$:
therefore, unless all the $\rai$ are identical,
a multi--component OM system is \textit{not} an OM system.
It is easy to prove that the DF of each component
is related to its density profile as
\begin{equation}\label{dirOM}
\rho=\int fd^3v=2\sqrt{8}\pi A(r,\ra) \int_0^{\psit}\sqrt{\psit-Q}f(Q)dQ,
\end{equation}
where
\begin{equation}\label{aOM}
A(r,\ra)=\frac{\ra^2}{\ra^2+r^2}.
\end{equation}
The radial dependence of the associated anisotropy parameter,
a quantity designed to measure
the differences between the tangential ($\sigt^2$)
and radial ($\sigr^2$) velocity dispersions, is
\begin{equation}\label{betaOM}
\beta(r)\equiv1-\frac{\sigt^2}{2\sigr^2}=\frac{r^2}{r^2+\ra^2}
\end{equation}
(Merritt 1985), so that the orbital distribution is isotropic at the center
and increasingly radially anisotropic with radius.
Note that consistency implies $\beta\leq1$.
With the introduction of the so--called ``augmented density''
\begin{equation}\label{rhoOM}
\varrho(r)\equiv\frac{\rho}{A(r,\ra)}=\left(1+\frac{r^2}{\ra^2}\right)\rho(r),
\end{equation}
it is possible to recast equation~(\ref{dirOM}) 
in a form suitable for Abel inversion, 
and after the differentiation one obtains
\begin{equation}\label{drhoOM}
\frac{d\varrho}{d\psit}=\sqrt{8}\pi\int_0^{\psit}\frac{f(Q)dQ}{\sqrt{\psit-Q}},
\end{equation}
where the function $\varrho$ is intended to be
expressed in terms of $\psit$, by the elimination of radius.
As first solved by Eddington (1916)
for the isotropic case in which $Q=\eps$,
equation~(\ref{drhoOM}) can be inverted as
\begin{eqnarray}\label{invOM}
f(Q)&=&\displaystyle\frac{1}{\sqrt{8}\pi^2}\frac{d}{dQ}\int_0^Q
\frac{d\varrho}{d\psit}\frac{d\psit}{\sqrt{Q-\psit}}\nonumber \\
&=&\displaystyle\frac{1}{\sqrt{8}\pi^2}\int _0^{Q}\frac{d^2\varrho}{
d\psit^2}\frac{d\psit}{\sqrt{Q-\psit}}
\end{eqnarray}
(Osipkov 1979), where the second identity above
holds for untruncated systems with
finite total mass.
Equation~(\ref{drhoOM}) is also of central importance
in the derivation of the CP92 necessary condition:

{\bf Theorem} [CP92, CM09] A \textit{necessary condition} (NC) for the
non-negativity of the DF of each density component $\rho$ 
in a multi--component OM system is
\begin{equation}\label{NCom}
\frac{d\varrho}{d\Psi}\geq 0,\quad 0\leq\Psi\leq\Psi(0),
\end{equation}
where $\varrho$ is the augmented density in equation~(\ref{rhoOM}), 
and $\Psi$ is the relative gravitational potential 
of the considered density component.
The NC can be rewritten in terms of the logarithmic density slope 
$\gamma(r)\equiv-d\ln\rho/d\ln r$ and of the anisotropy
parameter $\beta(r)$ as
\begin{equation}\label{cs}
\gamma(r)\geq 2\beta(r), \quad \forall r.
\end{equation}
In addition, a {\it weak sufficient condition} (WSC) 
for the non--negativity of each DF is
\begin{equation}\label{wsc}
\frac{d}{d\Psi}\left(\frac{d\varrho}{d\psit}\right)\geq
0, \quad 0\leq\Psi\leq\Psi(0) .
\end{equation}

\parn
{\bf Proof}: see CP92, C96, and CM09.
Here we just recall that the NC is obtained 
by assuming a positive $f(Q)$ in equation~(\ref{drhoOM}), 
while the WSC by requiring the positivity of the integrand
in the second equation~(\ref{invOM}), 
i.e. inequality~(\ref{wsc}) is nothing else
that a rewriting of $d^2\varrho/d\psit^2\geq0$.

Of particular relevance for the following discussion 
is inequality~(\ref{cs}), an unexpected extension of the ``central slope-anisotropy theorem'':

{\bf Theorem} [AE06] In all consistent \textit{constant anisotropy }
systems (with $\beta\leq1/2$) necessarily
\begin{equation}\label{ae}
\gamma(r)\geq 2\beta,\quad\forall r.
\end{equation}
Moreover, the same inequality holds asymptotically 
at the center (i.e., for $r\to0$) of any consistent spherical system
with \textit{generic anisotropy profile}.
\parn
{\bf Proof}: see Section 2.1.1 in AE06.

For completeness, we recall that systems with constant anisotropy 
are generated assuming a DF of the form
\begin{equation}\label{fCons}
f=J^{2\alpha}h(\eps),
\end{equation}
where $h(\eps)$ is a positive function,
and $\alpha>-1$ is a real number (see Section 3; see also Binney \& Tremaine 2008). 
In such models the anisotropy parameter is
\begin{equation}\label{betaalpha}
\beta(r)=-\alpha,
\end{equation}
so that for $\alpha>0$
they are characterized by tangential anisotropy,
while for $-1<\alpha<0$ the orbital anisotropy is radial.
The proof of identity~(\ref{betaalpha}) and 
the inversion formula analogous to~(\ref{invOM}) 
are not given here, being obtained 
as special cases of the Cuddeford systems described in the next Section.

\section{Consistency criteria for multi--component
generalized Cuddeford systems}
We begin this Section by recalling the main features
of the inversion for Cuddeford (1991) systems.
Then, in Section 3.2 the family of multi--component generalized Cuddeford systems
is introduced and the inversion formula obtained,
together with the associated consistency conditions.

\subsection{Cuddeford systems}
An interesting generalization of OM and constant anisotropy systems
was proposed by Cuddeford (1991; see also Ciotti 2000, Chapter 10)
assuming
\begin{equation}\label{f}
f=J^{2\alpha}h(Q),
\end{equation}
where $\alpha>-1$ is a real number
and $Q$ is defined as in equation~(\ref{fOM}):
isotropic models then correspond to $\alpha=0$
and $\ra\to\infty$.
Equation~(\ref{f}) can be used to describe both the OM models (for $\alpha=0$)
and the constant anisotropy models (for $\ra\to\infty$).
In particular, the anisotropy parameter 
takes now the simple form
\begin{equation}\label{beta}
\beta(r)=\frac{r^2-\alpha\ra^2}{r^2+\ra^2}
\end{equation}
(see equations~[\ref{sigma}]-[A5]).
Therefore, when $\alpha>0$ the anisotropy is tangential 
in the inner regions where $r<\sqrt{\alpha}\ra$, 
and radial for $r>\sqrt{\alpha}\ra$;
in the limit $\alpha\to\infty$, 
the orbital structure is fully tangentially anisotropic (i.e., $\beta\to-\infty$).
Instead, when $-1<\alpha<0$ the models are radially anisotropic everywhere,
independently of the value of $\ra$;
moreover, in the limit $\alpha\to-1$
equation~(\ref{beta}) gives $\beta\to1$, 
so that the velocity anisotropy is completely radial.

The DF of a Cuddeford system and its spatial density
are related as
\begin{equation}\label{cud}
\rho(r)=2\sqrt{8}\pi A(r,\alpha) \int_0^{\psit}(\psit-Q)^{\alpha+1/2}h(Q)dQ,
\end{equation}
where
\begin{equation}\label{aCud}
A(r,\alpha)=2^{\alpha-1}\sqrt\pi\frac{\Gamma(\alpha+1)}{\Gamma(\alpha+3/2)}\frac{r^{2\alpha}}{(1+r^2/\ra^2)^{\alpha+1}},
\end{equation}
and $\Gamma(x)=(x-1)!$ is the gamma function.
As expected, equation~(\ref{dirOM}) is reobtained 
for $\alpha=0$, while the convergence of the angular part of the integral 
over the velocity space requires $\alpha>-1$.
In analogy with the discussion in Section 2,
the augmented density
\begin{eqnarray}\label{rhoCud}
\varrho(r)&\equiv&\frac{\rho}{A(r,\alpha)}=\nonumber \\
&&\frac{2^{1-\alpha}}{\sqrt\pi}\frac{\Gamma(\alpha+3/2)}{\Gamma(\alpha+1)}
\left(1+\frac{r^2}{\ra^2}\right)^{\alpha+1}\frac{\rho(r)}{r^{2\alpha}}
\end{eqnarray}
is introduced, and a simple inversion formula, similar to equation~(\ref{invOM}),
permits to recover the DF from the density profile.
In fact, after
\begin{equation}\label{n}
m=\mbox{int}\left(\alpha+\frac{1}{2}\right)+1
\end{equation}
differentiations\footnote{$\mbox{int}(x)$ means the largest integer $\leq x$.
For example, $\mbox{int}(1/2)=0$ and so $m=1$ for OM models.} with respect to $\psit$, 
equation~(\ref{cud}) can be Abel inverted
(Cuddeford 1991).
In practice, one must perform enough differentiations
as to produce a negative exponent ($>-1$) in the power--law kernel
of integral~(\ref{cud}).

When $\alpha>-1$ (i.e., $m\geq0$)
but $\alpha$ is not half--integer,
\begin{eqnarray}\label{cudInv}
h(Q)&=&\frac{(-1)^{m+1}\cos{\alpha\pi}}{2\sqrt{8}\pi^2}
\frac{\Gamma(\alpha+3/2-m)}{\Gamma(\alpha+3/2)}\times\nonumber\\
&&\frac{d}{dQ}\int_0^{Q}\frac{d^{m}\varrho}{d\psit^{m}}\frac{d\psit}{(Q-\psit)^{\alpha+3/2-m}}\nonumber\\
&=&\frac{(-1)^{m+1}\cos{\alpha\pi}}{2\sqrt{8}\pi^2}
\frac{\Gamma(\alpha+3/2-m)}{\Gamma(\alpha+3/2)}\times\nonumber\\
&&\int_0^{\psit}\frac{d^{m+1}\varrho}{d\psit^{m+1}}\frac{d\psit}{(Q-\psit)^{\alpha+3/2-m}},
\end{eqnarray}
where the last identity holds for untruncated systems
with finite total mass, and the OM inversion formula~(\ref{invOM}) 
is reobtained for $\alpha=0$.

When $\alpha$ is half--integer, i.e.
$\alpha=m-3/2$ and $m=1,2,...$,
the solution of the Volterra equation~(\ref{cud}) is given by
\begin{equation}\label{cudInvInt}
h(Q)=\frac{1}{2\sqrt{8}\pi (m-1)!}
\left[\frac{d^m\varrho}{d\psit^m}\right]_{\psit=Q},
\end{equation}
and the DF is recovered analytically avoiding integration\footnote{In
equation~(30) of Cuddeford (1991) the $(m-1)!$ at the denominator is missing.
See also equations~(49) and (51) of Baes \& Dejonghe (2002). }.

\subsection{The consistency criteria and the density slope--anisotropy inequality}
As we now show, the inversion formulae~(\ref{cudInv}) and~(\ref{cudInvInt})
still hold for the more general case of 
multi--component, generalized Cuddeford systems,
in which  the DF associated with \textit{each} density component 
is made by the sum of an arbitrary number 
of Cuddeford DFs with arbitrary positive weights $w_i$
and possibly different anisotropy radii $\rai$
(but same $h$ function and angular momentum exponent), i.e.
\begin{equation}\label{sumCud}
f=J^{2\alpha}\sum_i w_i h(Q_i),\quad Q_i=\eps-\frac{J^2}{2\rai^2}.
\end{equation}
The different density components
of a multi--component generalized Cuddeford system
will have, in general, a different value of $\alpha$
and a different function $h(Q)$.
As should be clear, all the results presented in Sections 2 and 3.1
hold as special cases of the following treatment.

Of course, the orbital anisotropy distribution characteristic 
of DF~(\ref{sumCud}) is \textit{not} a Cuddeford one:
as shown in the Appendix, the anisotropy function $\beta(r)$
of each density component is given by 
\begin{equation}\label{betaCud}
\beta(r)=1-(\alpha+1)\frac{\sum_i w_i/(1+r^2/\rai^2)^{\alpha+2}}{\sum_i w_i/(1+r^2/\rai^2)^{\alpha+1}}.
\end{equation}
Quite general anisotropy profiles
can be obtained by specific choices of the weights $w_i$,
the anisotropy radii $\rai$, and the exponent $\alpha$.
However, near the center $\beta(r)\sim-\alpha$,
and $\beta(r)\sim1$ for $r\to\infty$,
independently of the specific values of $w_i$ and $\rai$.

We now show that an Abel inversion formula 
identical to equations~(\ref{cudInv})-(\ref{cudInvInt})
can be found for a DF of the family~(\ref{sumCud}).
In fact, it is immediate to verify that equation~(\ref{cud}) still holds, 
where now the radial function is 
\begin{equation}\label{aCudsum}
A(r,\alpha)=\frac{\sqrt\pi}{2}\frac{\Gamma(\alpha+1)}{\Gamma(\alpha+3/2)}\sum_i\frac{ w_i
r^{2\alpha}}{(1+r^2/\rai^2)^{\alpha+1}},
\end{equation}
and so, once the new augmented density $\varrho=\rho/A$ is defined,
the function $h$ in equation~(\ref{sumCud})
can in principle be recovered.
Therefore, it is obvious that the same arguments 
used to derive the necessary and sufficient conditions
for consistency of OM models can be repeated also 
for each density component of
multi--component generalized Cuddeford systems.
However, as $m$ differentiations with respect to $\psit$
must be performed on the integral~(\ref{cud}) before the inversion, 
we now obtain $m$ necessary conditions and a sufficient condition.
Surprisingly, as in the case of OM models, we found that 
the first of the necessary conditions for consistency
can be rewritten as the density slope--anisotropy theorem
which must hold at every radius.
These results are summarized in the following

{\bf Theorem} Each density component in a consistent
multi--component generalized Cuddeford system 
with $\alpha$ not half--integer 
obeys $m$ necessary conditions ($\rm{NC}_k$):
\begin{equation}\label{nc}
\frac{d^k\varrho}{d\psit^k}\geq0, \quad k=1,2,... m,
\end{equation}
where $m$ is given by equation~(\ref{n}).
In particular, the $\rm{NC}_1$ can be rewritten as the 
density slope--anisotropy inequality
\begin{equation}\label{csCud}
\gamma(r)\geq 2\beta(r),\qquad\forall r.
\end{equation}
Moreover, a sufficient condition for the non--negativity of the DF 
of each component is 
\begin{equation}\label{sc}
\frac{d^{m+1}\varrho}{d\psit^{m+1}}\geq0.
\end{equation}

\parn
{\bf Proof}: 
A proof of the $m$ necessary conditions~(\ref{nc}) is obtained
by repeated differentiation of the augmented density $\varrho$
(see equations~[\ref{cud}]-[\ref{rhoCud}], 
where now $A(r,\alpha)$ is given by equation~[\ref{aCudsum}])
with respect to the total potential $\psit$,
and by the assumption that $h(Q)$ is a positive function.
The sufficient condition~(\ref{sc}) is derived
just by imposing the positivity of the integrand
in the second of identities~(\ref{cudInv}).
Finally, we refer to the Appendix 
for a proof of inequality~(\ref{csCud}).
\medskip

Of course, in the special cases of $\alpha=m-3/2$ 
and $m=1,2...$, equation~(\ref{cudInvInt})
provides, in addition to the $m-1$ necessary conditions~(\ref{nc}),
the necessary and sufficient condition
for consistency of the specific component, i.e.
\begin{equation}\label{nsc}
\frac{d^m\varrho}{d\psit^m}\geq0.
\end{equation}
As expected, the NC and the WSC derived in CP92
are reobtained as special cases of the new theorem
for $\alpha=0$ (i.e. $m=1$).
In applications, as those that will be presented in Section 4,
it is useful to express equations~(\ref{nc}) and~(\ref{sc})
in terms of the radius.
From the relation $d\Psi_{\rm T}/dr=-GM_{\rm T}(r)/r^2$,
where $M_{\rm T}(r)$ is the total mass enclosed by the radius $r$,
the $\rm{NC}_1$ can be expressed as 
\begin{equation}\label{nc1}
\frac{d\varrho}{d r}\leq0,
\end{equation}
which is a second alternative formulation of 
the density slope--anisotropy theorem
in addition to equation~(\ref{csCud}).
Following the same approach, the $\rm{NC}_2$ can be also expressed as
\begin{equation}\label{nc2}
\frac{d}{d r}\left[\frac{r^2}{M_{\rm T}(r)}\frac{d\varrho}{d r}\right]\geq0,
\end{equation}
and so on, with the sign of the $\rm{NC}_k$ inequality 
alternating with increasing $k$.
Finally, note that $\rm{NC}_1$ is the sole condition
in which only the augmented density profile of the specific density component appears,
while in the higher order $\rm{NC}_k$
the total mass profile $M_{\rm T}(r)$ is also involved.

\section{Some illustrative cases}
In the previous Section we derived
the family of necessary and sufficient conditions
for phase--space consistency of each density component of generalized Cuddeford systems,
and we showed that the density slope--anisotropy inequality
holds at every radius.

We now present a simple application
of the new consistency criteria,
and we address two natural questions concerning 
phase--space consistency of Cuddeford systems.
The first is related to the fact that for $\alpha\geq1/2$ we have,
at variance with the OM case ($\alpha=0$),
more than one necessary condition
for the non--negativity of the DF.
Which necessary condition is stronger? 
Or, more quantitatively, which of the $\rm{NC}_k$ gives a consistency limit
closer to the true one (that would be derived from the DF)?
The second question is: for a given density profile,
what is the effect of tangential anisotropy on consistency?
Will the minimum anisotropy radius
increase or decrease at increasing $\alpha$,
i.e. at increasing tangential anisotropy?
The set of necessary conditions
and the dependence of their number on $\alpha$ 
through equation~(\ref{n})
suggest a simple approach 
to address the two issues above.
Consider an assigned density profile,
representing a component in a multi--component
generalized Cuddeford system: 
what is the behaviour of the consistency region in parameter space 
at increasing $\alpha$?
At increasing $\alpha$ the number of necessary conditions increases:
of course each additional necessary condition
can only reduce the consistency region in the parameter space.
In addition, when $\alpha$ increases so that $m$
given by equation~(\ref{n}) increases by 1,
the former sufficient condition $\rm{NC}_{m+1}$ 
becomes the last of the necessary conditions
for the new model.

We now illustrate the procedure, and discuss the two questions presented above,
by investigating the phase--space consistency of the widely used $\gamma$--models
(Dehnen 1993, Tremaine et al. 1994; see equation~(\ref{gammamod}) below).
We do this in the most simplified form,
i.e. in the case of a one--component Cuddeford system;
in other words, in equation~(\ref{sumCud}) we restrict to $i=1$.
The detailed study of $\gamma=0$, $\gamma=1$ (Hernquist 1990), 
and $\gamma=2$ (Jaffe 1983) models
will also allow us to explore the combined effect of the inner density slope
and of tangential anisotropy (Section 4.1),
while the additional role played by the external density slope
will be discussed in Section 4.2 by using one--component $n$--$\gamma$ models with Cuddeford anisotropy
(see equation~(\ref{ngamma}) below).
We recall that a consistency analysis of OM anisotropic  $n$--$\gamma$ models
was done in CM09.

\subsection{The one--component $\gamma$--models} 
\begin{figure}
\includegraphics[scale=0.4]{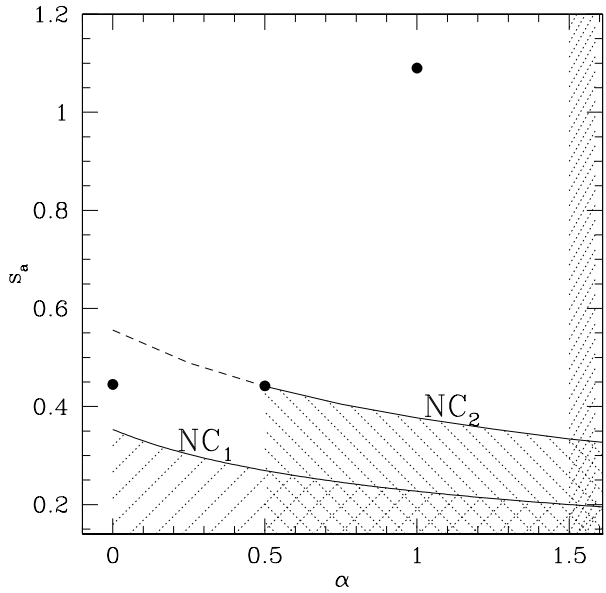}
\caption{Consistency limits on the normalized anisotropy radius $\sa=\ra/\rc$ 
for the one--component $\gamma=0$ model with Cuddeford anisotropy. 
The solid curves mark the limits 
imposed by $\rm{NC}_1$ and  $\rm{NC}_2$: 
models with the pair ($\alpha$,$\sa$) in the shaded regions
are certainly inconsistent; models above the dashed curve
(where $\rm{NC}_2$ is the sufficient condition~[\ref{sc}])
are certainly consistent. Solid dots are the true lower limits for $\sa$
derived from the DF. Recall that $\alpha=0$ refers to the OM model.
No consistent models exist for $\alpha\geq3/2$.}
\label{fig:g0}
\end{figure}

\begin{figure}
\includegraphics[scale=0.4]{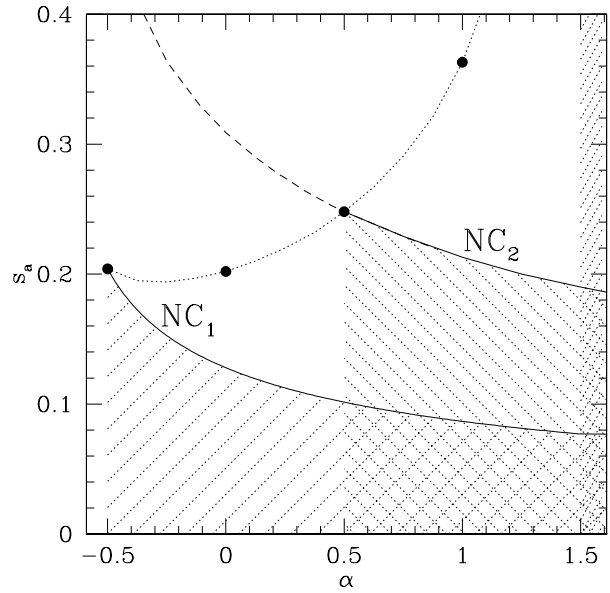}
\caption{Consistency limits on the normalized anisotropy radius $\sa=\ra/\rc$ 
for the one--component Hernquist ($\gamma=1$) model with Cuddeford anisotropy.
Different curves have the same meaning as in Fig.~\ref{fig:g0}.
The dotted line connecting the solid dots has been obtained from Table 1 
in Baes \& Dejonghe (2002). }
\label{fig:her}
\end{figure}

\begin{figure}
\includegraphics[scale=0.4]{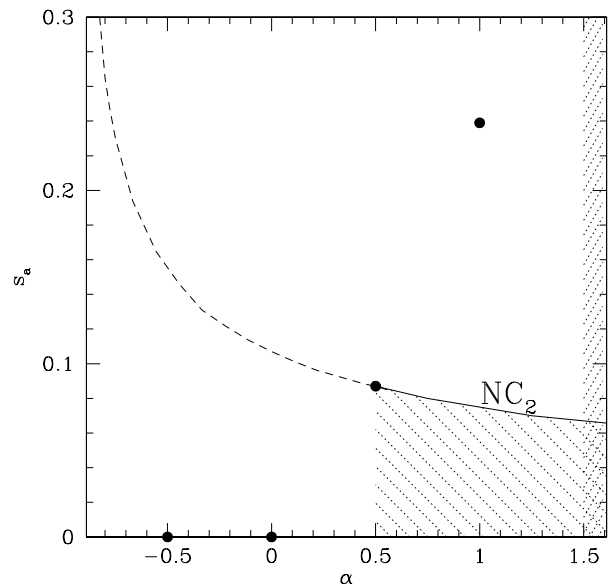}
\caption{Consistency limits on the normalized anisotropy radius $\sa=\ra/\rc$ 
for the one--component Jaffe ($\gamma=2$) model with Cuddeford anisotropy. 
Different curves have the same meaning as in Fig.~\ref{fig:g0}.
Note that the $\rm{NC}_1$ coincides with the $x$--axis.}
\label{fig:jaf}
\end{figure}

We start by considering the general $\gamma$--model,
whose dimensionless density profile and mass enclosed inside radius $r$ are given by
\begin{eqnarray}\label{gammamod}
 &\rho(r)&=\frac{1}{s^\gamma(1+s)^{4-\gamma}},\\
 &M(r)&=\left(\frac{s}{1+s}\right)^{3-\gamma},\qquad0\leq\gamma<3,
\end{eqnarray}
where $s\equiv r/\rc$ is the radius normalized
to the ``core'' radius $\rc$. 
It is trivial to show that the consistency properties
of one--component models are independent 
of the mass and density normalization scales.

In the following we will study the $\rm{NC}_k$ functions
by using their radial formulation (equations~[\ref{nc1}]-[\ref{nc2}]
with the augmented density of equation~[\ref{rhoCud}]).
Indeed, in common situations 
the elimination of the radius from the density profile 
in favour of the gravitational potential,
needed to evaluate equation~(\ref{nc}),
is not feasible. For this reason we prefer to study the consistency conditions
by using their radial expressions, as this procedure is always viable, 
for whatever density profile expressed as a function of radius.

We begin by noticing that from AE06 theorem we already know
that $\alpha\geq-\gamma/2$ is required at the center of the density distribution~(31);
this condition must be combined with $\alpha>-1$ (see Section 3.1).
Once the appropriate augmented density is defined, the radial $\rm{NC}_1$ 
for Cuddeford anisotropic $\gamma$--models read with equation~(\ref{rhoCud}) reduces to
\begin{equation}\label{ncgamma}
\sa^2[2s(2+\alpha)+2\alpha+\gamma]+s^2(2s+\gamma-2)\geq0, \qquad\forall s,
\end{equation}
thus establishing a relation between $\alpha$ 
and the normalized anisotropy radius $\sa\equiv \ra/\rc$.

As expected, for $s=0$ the inequality above 
reduces to the AE06 limitation.
However, as the condition~(\ref{ncgamma}) 
must hold over the entire radial range,
we can now derive limitations 
on the minimum allowed anisotropy radius $\sa$ 
as a function of $\gamma$ and $\alpha$.
The general formula is simple but here we prefer 
to focus on the special cases of $\gamma=0,1,$ and 2.
The $\rm{NC}_1$ is represented by a solid curve
in Figs~\ref{fig:g0}, \ref{fig:her}, \ref{fig:jaf} respectively
for the $\gamma=0,1$, and $2$ models;
of course, while $\alpha$ is restricted to positive values when considering the $\gamma=0$ case,
the $\alpha$ axis begins at $\alpha=-0.5$ for $\gamma=1$ models, 
and finally the AE06 limitation in the $\gamma=2$ case is $\alpha>-1$ (coincident with the value required
by convergence of the integral in equation~[\ref{cud}]).

In the three figures, all points below the solid $\rm{NC}_1$ curve correspond to unphysical models,
while points above may represent consistent models.
The solid dots are the true lower limits on $\sa$
determined by direct inspection of the DF
for representative values of $\alpha$:
the $\gamma=1$ case was already given by Baes \& Dejonghe (2002).
Note that for the Jaffe model (Fig.~\ref{fig:jaf}), 
the $\rm{NC}_1$ actually coincides with the abscissae axis,
i.e. it is satisfied for all values of $\alpha>-1$, 
independently of the value of $\sa$.
In the case of $\gamma=1$ and $\gamma=2$ models,
when $\alpha=-1/2$ equation~(\ref{cudInvInt}) provides the DF,
and so the true critical anisotropy radius $\sa$
can be easily determined (black dots in Fig.~\ref{fig:her} and \ref{fig:jaf}). 
Of course, these values coincide with those 
obtained from the $\rm{NC}_1$ for $\alpha=-1/2$,
as should be clear from the discussion in Section 3.
The solid dots at $\alpha=0$ represent instead the DF--derived lower limit
for the minimum anisotropy radius 
for the corresponding OM models (see, e.g., C96).
As the $\rm{NC}_1$ for $-0.5<\alpha<0.5$ is 
just a necessary condition for consistency,
while the $\rm{NC}_2$ provides a \textit{sufficient} condition
for consistency (i.e. all points above the $\rm{NC}_2$ dashed curve
correspond to consistent models), 
it is not surprising that for all the three density models 
the DF--derived limit on the anisotropy radius
in the OM case is contained in the region
delimited by the $\rm{NC}_1$ and the $\rm{NC}_2$
(see C99, Table 1).

As we increase $\alpha$, when we reach the value $\alpha=1/2$
the $\rm{NC}_2$ function becomes the model DF, 
and so the DF--derived lower limit, represented with a black dot,
coincides again with the critical curve.

For $1/2<\alpha<3/2$, the $\rm{NC}_2$ becomes a new necessary condition,
and therefore all points in the shaded area below the solid $\rm{NC}_2$ curves
in the three figures correspond to unphysical models.
Note how the $\rm{NC}_2$ provides
more stringent limits than the $\rm{NC}_1$.
Consistently with the nature of the $\rm{NC}_2$, 
the black dots representing the limits on $\sa$ derived from the DF 
for $\alpha=1$ lie above the $\rm{NC}_2$ curve.
Of course, in this range of values of $\alpha$ the $\rm{NC}_3$
is the sufficient condition for phase--space consistency.
However, an asymptotic expansion of the $\rm{NC}_3$ for $s\to\infty$ 
easily shows that this condition is violated,
independently of the value of $\sa$ and $\alpha$.
This fact poses no problem in the range $1/2<\alpha<3/2$,
as $\rm{NC}_3$ is a sufficient condition there, 
but as soon as $\alpha$ becomes larger than $3/2$ 
the $\rm{NC}_3$ becomes necessary,
and the whole family of $\gamma$--models
with Cuddeford anisotropy becomes inconsistent.
We note that the $\alpha=3/2$ limitation was already determined
by Baes \& Dejonghe (2002) for Hernquist models with Cuddeford anisotropy.
Quite surprisingly, by using the $\rm{NC}_3$ 
we found that the limitation $\alpha<3/2$ 
holds for the entire family of $\gamma$--models, 
no matter which value of $\gamma$ is considered.
The reason is due to the fact that the \textit{external} density slope
of $\gamma$--models is 4, independently of the value of $\gamma$.
Thus, while the lower limit on $\alpha$ is due 
to the central density slope,
the external density slope limits the amount 
of tangential anisotropy that can be supported by the models.
This indication is very interesting, 
because it means that the external regions
(where anisotropy is almost completely radial,
see equation~[\ref{beta}])
are able to affect the inner dynamics.
We will discuss such issue in the next Section 4.2.

As a final remark, we note that 
a comparison of Figs~\ref{fig:g0}, \ref{fig:her}, and \ref{fig:jaf}
confirms qualitatively the trend already found in Carollo et al. (1995),
C96 and C99 for one--component OM models.
In practice, at fixed $\alpha$ 
the minimum anisotropy radius increases 
at decreasing inner density slope $\gamma$,
i.e. centrally flatter density profiles
are less able to sustain radial anisotropy
than steeper density profiles,
even in presence of a central tangential anisotropy.
This is shown by the smaller and smaller shaded areas at 
fixed $\alpha$ and increasing $\gamma$, and by the corresponding
smaller exact values of the minimum $\sa$ indicated by the solid dots.

\subsection{The effect of the external density slope}
As we have seen, no Cuddeford anisotropic $\gamma$--model
exists for $\alpha\geq3/2$,
and this independently of the value of $\gamma$.
The fact that the critical upper limit
of $\alpha$ is independent of $\gamma$
is a clear indication of the importance of the slope
of the outer density profile on the central anisotropy.
However, being the limit imposed on $\alpha$,
it implies that the central regions
are the ones affected.
A hint to understand this phenomenon
is given by inspection of the three figures: in fact, note how
the consistency region in the ($\alpha$,$\sa$) space
reduces at increasing $\alpha$,
in the sense that for increasing $\alpha$
the minimum value of $\sa$ increases.
This means that when the central regions are forced
to be more and more tangentially anisotropic,
the external regions (where $\sa$ determines
the amount of radial anisotropy) must be more and more isotropic.
Therefore, we conclude that the origin
of the inconsistency at high $\alpha$
is a combination of the forced tangential anisotropy
and the radial orbits arriving 
from the external regions of the system.

To better understand this behaviour, 
we now consider the one--component,
Cuddeford anisotropic $n$--$\gamma$ models, 
whose normalized density profile is given by
\begin{equation}\label{ngamma}
 \rho=\frac{1}{s^\gamma(1+s)^{n-\gamma}},\qquad 0\leq\gamma<3,\qquad n>3
\end{equation}
(see CM09).
If our previous argument is correct,
then the maximum value of $\alpha$
should increase at increasing $n$,
as less and less mass is contained
outside the core radius at increasing $n$,
so that less and less radial orbits can affect the inner regions.
Unfortunately, for generic (non integer) values of $n$
the mass contained within $r$ cannot be expressed
in terms of elementary functions.
However, it is possible to perform 
an asymptotic analysis at large radii of the $\rm{NC}_k$
(with some care,  as differentiation of asymptotic relations 
is usually not permitted, e.g. see Bender \& Orszag 1978).
The appropriate way to perform the analysis in this case 
is to use the $\rm{NC}_k$ formulated 
in terms of the potential (equation~[\ref{nc}]),
and to adopt the asymptotic expansion for the relative potential 
$\Psi=1/s+O(1/s^2)$.
Following this approach it can be proved that, 
independently of the value of the inner density slope $\gamma$,
the critical value of $\alpha$ increases with $n$:
for example, when $n=5$ it is required that $\alpha<5/2$,
when $n=6$ that $\alpha<7/2$, and so on.
This confirms the previous conjecture.

\section{Discussion and conclusions}
In a natural extension of previous investigations
(CP92, C96, C99, AE06, CM09),
we searched for phase--space consistency criteria
for multi--component spherical systems.
We found that inversion formulae and necessary and sufficient conditions
for consistency can actually be derived 
for multi--component generalized Cuddeford systems.
Such systems contain as very special cases 
OM, constant anisotropy, and Cuddeford models.
The main results of our study can be summarized as follows:

\begin{enumerate}

\item New phase--space consistency criteria,
i.e. necessary and sufficient conditions for the DF non--negativity, 
are derived for multi--component, generalized Cuddeford systems.
At variance with the simpler case of OM models,
the presence of tangential anisotropy
leads to a \textit{family} of necessary conditions,
that can be written as simple inequalities
involving repeated differentiations
of the augmented density expressed
as a function of the total potential.

\item It is shown that the first of the necessary conditions
for consistency can be reformulated as the density slope--anisotropy theorem,
which therefore is proved to hold not only at the center 
\textit{but also at all radii for each density component
of multi--component generalized Cuddeford models}.

\item The first necessary condition
is the only condition independent
of the other density components of the model.
All the other (more stringent) conditions
depend on the total density distribution of the model.

\item All the conditions can be reformulated
in term of the radius, so that they can be tested
also for models in which the total potential 
cannot be expressed by using elementary functions,
or when the radius cannot be eliminated in favour of the potential.

\item The new phase--space consistency criteria
are applied to one--component $\gamma$--models
with Cuddeford anisotropy.
It is found that for increasing tangential anisotropy
in the central regions the minimum anisotropy radius for consistency
increases, i.e. the external regions
must be less and less radially anisotropic.
No consistent $\gamma$--models
exist for $\alpha\geq3/2$,
independently of the value of the central density slope $\gamma$.
Baes \& Dejonghe (2002) already found this limitation
by direct inspection of the DF of Hernquist models with Cuddeford anisotropy.

\item To investigate the combined effect
of the outer radial and inner tangential anisotropy,
we performed an asymptotic analysis
of one--component $n$--$\gamma$ models
with Cuddeford anisotropy.
We found that a steepening of the external density slope
allows larger values of the central tangential anisotropy,
independently of the value of the central density slope $\gamma$,
thus confirming the hypothesis of a dynamical interplay
between the two regions of the models,
and supporting the interpretation 
that Baes \& Dejonghe (2002) proposed for Hernquist models.
\end{enumerate}

We notice that one of the major results of this study seems to be 
the generality of the density slope--anisotropy relation 
$\gamma(r)\geq2\beta(r)$.
It is natural to ask 
whether such density slope--anisotropy relation 
is even more general, i.e. it is an inequality necessarily obeyed
by generic spherically symmetric, two--integrals systems 
with positive DF.
At this stage we do not have a proof of this conjecture,
but we are not aware of any counter--example.
Actually, we have additional evidences 
supporting this conjecture:
for example Michele Trenti kindly provided us 
with a large set of numerically computed $f_{\nu}$ models (Bertin \& Trenti 2003),
and all of them, without exception,
satisfy the inequality $\gamma(r)\geq2\beta(r)$ at all radii.
Moreover, it is trivial to show
that spherical systems in which the density
can be written as $\rho=A(r)f(\Psi)$,
with $f$ monotonically increasing function of $\Psi$,
all obey to $\gamma(r)\geq2\beta(r)$
when supported by a positive DF
(see also the comment after equation~[\ref{last}]).
We stress that these models do not belong 
to the family of generalized Cuddeford systems.
Other distributions of orbital anisotropy
that are not of the Cuddeford family
(even though they could be approximated by specific choices
of generalized Cuddeford distributions)
have been reported by Mamon \& Lokas (2005),
Wojtak et al. (2008, who went further to show 
that also the DF was not of the OM or Cuddeford forms), Ascasibar et al. (2008)
from the analysis of halos in cosmological simulations,
or proposed in terms of specific DFs (e.g., Gerhard 1991;
Louis 1993; Cuddeford \& Louis 1995):
it would be interesting to check 
the $\gamma(r)\geq2\beta(r)$ inequality in these systems.
In any case, we note that numerical simulations
are known to produce correlations
between $\beta$ and $\gamma$ 
(e.g., see Hansen \& Moore 2006, Mamon et al. 2006).
We finally conclude by noticing that,
if the inequality $\gamma(r)\geq2\beta(r)$ is universal (for spherical systems),
then the so--called mass-anisotropy degeneracy
could be less severe than what is sometimes feared,
as orbital anisotropy would be in some sense
controlled by the local density slope 
of the stellar distribution in galaxies (in the inner regions where $\gamma\leq2$).
This could be an important constraint 
in observational works.

\section*{Acknowledgments} We thank Michele Trenti
for having tested the density slope--anisotropy inequality~(\ref{csCud}) 
for a large set of $f_{\nu}$ models,
and the referee, Gary Mamon, for a careful reading
and for very helpful comments that improved the presentation.

\appendix
\section{The density slope--anisotropy inequality 
for generalized Cuddeford systems}

The radial and tangential velocity dispersion profiles
of each density component of a multi--component generalized Cuddeford system
are given by
\begin{eqnarray}\label{sigma}
\rho\sigma^2_r &=& \int f v^2_{\rm r} d^3v\nonumber\\
&=&4\pi B(r,\alpha) \int_0^{\psit}[2(\psit-Q)]^{\alpha+3/2}h(Q)dQ, \\
\rho\sigma^2_t &=& \int f v^2_{\rm t} d^3v\nonumber\\ 
&=& 4\pi C(r,\alpha) \int_0^{\psit}\left[2(\psit-Q)\right]^{\alpha+3/2}h(Q) dQ,
\end{eqnarray}
where
\begin{eqnarray}\label{BC}
B(r,\alpha)&\equiv&\frac{\sqrt\pi}{4}\frac{\Gamma(\alpha+1)}{\Gamma(\alpha+5/2)}\sum_i\frac{ w_i r^{2\alpha} }{(1+r^2/\rai^2)^{\alpha+1}},\\
C(r,\alpha)&\equiv&\frac{\sqrt\pi}{2}\frac{\Gamma(\alpha+2)}{\Gamma(\alpha+5/2)}\sum_i\frac{ w_i r^{2\alpha}}{(1+r^2/\rai^2)^{\alpha+2}}.
\end{eqnarray}
Then, from equation~(\ref{betaOM}) one has
\begin{equation}\label{betaCudsum} 
\beta(r)=1-\frac{C(r,\alpha)}{2B(r,\alpha)},
\end{equation}
and simple algebra proves equation~(\ref{betaCud}).

We now show that the inequality $\gamma(r)\geq2\beta(r)$ holds
at all radii in each density component
of consistent generalized Cuddeford systems.
First, we relate the logarithmic density slope
$\gamma(r)\equiv-d\ln\rho/d\ln r$ to the $\rm{NC}_1$~(\ref{nc1}) as follows:
\begin{equation}
0\geq\frac{d\varrho}{dr}=\frac{d}{dr}\frac{\rho}{A}
=\frac{\rho}{rA}\frac{d\ln(\rho/A)}{d\ln r},
\end{equation}
so that the $\rm{NC}_1$ can be simply rewritten as
\begin{equation}\label{genCs}
\gamma(r)\geq-\frac{d\ln A}{d\ln r}.
\end{equation}
In other words, all consistent generalized Cuddeford systems
satisfy equation~(\ref{genCs}) at each radius.
Now it is easy to verify that the functions $A$, $B$, and $C$,
given in equations~(\ref{aCudsum}), (\ref{BC}), and (A4),
satisfy the identity
\begin{equation}\label{nc4nc}
-\frac{d\ln A}{d\ln r}=2\left[1-\frac{C(r,\alpha)}{2B(r,\alpha)}\right]=2\beta
\end{equation}
for arbitrary $\alpha$, $w_i$ and $\rai$,
so that equations~(\ref{genCs}) and~(\ref{nc4nc}) show 
that the inequality $\gamma(r)\geq2\beta(r)$ 
is just another way to express the $\rm{NC}_1$.
Identity~(\ref{nc4nc}) can be proved by elementary algebra:
\begin{eqnarray}\label{last}
&&-\frac{d\ln A}{d\ln r}=-2\alpha\\
&&+2(\alpha+1)\sum_i\frac{ w_i r^2 +w_i\rai^2-w_i\rai^2 }{\rai^2(1+r^2/\rai^2)^{\alpha+2}}\left[\sum_i\frac{ w_i }{(1+r^2/\rai^2)^{\alpha+1}}\right]^{-1},\nonumber
\end{eqnarray}
where in each term of the sum we added and subtracted $w_i\rai^2$.
Simplification and comparison with equation~(\ref{betaCud}) conclude the proof.

We note that identity~(\ref{nc4nc}) is actually a special case
of a more general result reported in Baes \& Dejonghe (2002)
and Baes \& van Hese (2007),
which holds for \textit{all} spherical systems whose DF,
after integration over velocity space,
leads to the factorization $\rho(r)=A(r)f(\Psi)$;
for such systems (that also include our generalized Cuddeford systems)
it can be proved that $2\beta(r)=-d \ln A/d\ln r$.


\end{document}